\documentclass[preprint,aps]{revtex4}
\usepackage{amstext}
\usepackage{graphicx}
\usepackage{psfrag}

\begin{document}
\title{Spin and Charge Dynamics in a Renormalised Perturbation  Theory}
\author{A.C. Hewson}\email{a.hewson@imperial.ac.uk}
\affiliation{Dept. of Mathematics, Imperial College, London SW7 2BZ.}

\date{\today}

\begin{abstract}
We calculate the  spin and charge dynamical susceptibilities of a strongly correlated
impurity model in a renormalised perturbation theory.
 The irreducible for vertices for the quasiparticle scattering
 are deduced from the renormalised parameters, which have been calculated by fitting
of the low-lying levels of a numerical renormalization
group (NRG) calculation to those of an effective Anderson model.
The susceptibilities are asymptotically exact in the low frequency
limit and satisfy the Korringa-Shiba relation. By comparing the results with
those calculated from a direct NRG calculation, we show that the renormalised
perturbation theory (RPT) description gives a very good description of 
 spin dynamics for all values of the local interaction  U,
 not only at low frequencies but over the whole
relevant frequency range. 

In the presence of a magnetic field the approach can be generalised using
field dependent renormalized parameters to calculate  the transverse and
parallel spin dynamic susceptibilities. The RPT results give accurate results for
the 
 spin dynamics over the whole frequency range and for all values of the
 magnetic field. 

   \end{abstract}
\maketitle
\pagestyle{empty}

\section{Introduction}
The scattering of electrons with  low energy  spin excitations is responsible
for the anomalous behaviour of many systems which are described as being
strongly correlated. In magnetic impurity systems this scattering is enhanced
leading to 
the Kondo effect and the many-body resonance (Kondo resonance) which develops
at low temperatures at the Fermi level. It is also  the origin
of the large mass enhancements in heavy fermion systems, and the strong
renormalisation of the energy bands at the Fermi level. At a quantum critical
point, where  the critical temperature for magnetic ordering is reduced
to zero by pressure or alloying, it is the local low energy fluctuations
which cause the breakdown of Fermi liquid theory. In the cuprate high
temperature superconductors, it is the scattering with the spin fluctuations, after
the antiferromagnetic order has been destroyed through doping, which is the
likely cause of  the
anomalous  electronic behaviour in the normal state. The exchange
of the magnetic fluctuations between the conduction electrons may be the
origin of the electron attraction leading to superconductivity. It is
important, therefore, to understand the nature of these fluctuations
in the absence of any long range magnetic order. Here, we study the spin
dynamics in a strongly correlated local model, the impurity Anderson model
\cite{And61}, and calculate 
local spin dynamics using a renormalised perturbation theory expansion. We start
by describing the model and the technical
aspects of the approach in this section of the paper, and later present results for various
parameter regimes, both with and without an applied magnetic field.\\
 
 The  Anderson model   has the form,
\begin{equation} H_{\rm AM}=\sum\sb {\sigma}\epsilon\sb {d}
d\sp {\dagger}\sb {\sigma}
d\sp {}\sb {\sigma}+
Un\sb {d,\uparrow}n\sb {d,\downarrow}
 +\sum\sb {{ k},\sigma}( V\sb { k}d\sp {\dagger}\sb {\sigma}
c\sp {}\sb {{ k},\sigma}+ V\sb { k}\sp *c\sp {\dagger}\sb {{
k},\sigma}d\sp {}\sb {\sigma})+\sum\sb {{
k},\sigma}\epsilon\sb {{ k},\sigma}c\sp {\dagger}\sb {{ k},\sigma}
c\sp {}\sb {{
k},\sigma},\label{ham}\end{equation}
where $\epsilon_d$ is the energy of the impurity level, $U$ is the interaction at the impurity site,
 and $V_{k}$ the hybridization matrix element to a band of conduction electrons with
energy $\epsilon_k$. In the wide band limit the hybridization weighted density of states,
 $\Delta(\omega)=\pi\sum_{k}|V_k|^2\delta(\omega-\epsilon_k)$, can be taken as
 a constant $\Delta$. With this assumption the impurity one-electron with spin  $\sigma$
 Green's
 function is given by
\begin{equation}G_{d\sigma}(\omega)={1\over\omega+i\Delta-\epsilon_d-\Sigma_\sigma(\omega)}\end{equation}
where $\Sigma_\sigma(\omega)$ is the self-energy.\\

We have demonstrated  in earlier work that the low energy fixed point of this model \cite{Wil75,KWW80} can be interpreted in terms of the same model but with renormalised
parameters \cite{Hew93,Hew01}. The renormalised model
  takes the form,
\begin{equation} 
\tilde H_{\rm AM}=\sum\sb {\sigma}\tilde\epsilon\sb {\mathrm{d}}
d\sp {\dagger}\sb {\sigma}
d\sp {}\sb {\sigma}+
\tilde U : n\sb {{\rm d},\uparrow}n\sb {{\rm d},\downarrow} :
+\sum\sb {{ k},\sigma}(\tilde V\sb { k}d\sp {\dagger}\sb {\sigma}
c\sp {}\sb {{ k},\sigma}+\tilde V\sb { k}\sp *c\sp {\dagger}\sb {{
k},\sigma}d\sp {}\sb {\sigma})+\sum\sb {{
k},\sigma}\epsilon\sb {{ k},\sigma}c\sp {\dagger}\sb {{ k},\sigma}
c\sp {}\sb {{
k},\sigma},\label{rham}
\end{equation}
where  the colon brackets indicate that the expression within them must be
normal-ordered. 
The three parameters which specify this renormalised model are 
$\tilde\epsilon_d$, $\tilde\Delta$ and $\tilde U$, defined by
\begin{equation}\tilde\epsilon_d=z(\epsilon_d+\Sigma(0)),\quad\tilde\Delta =z\Delta,\quad \tilde U=z^2\Gamma_{\uparrow\downarrow}(0,0,0,0),
\label{ren}\end{equation}
where $z$ is given by
$z={1/{(1-\Sigma'(0))}}$ and
$\Gamma_{\uparrow\downarrow}(\omega_1,\omega_2,\omega_3,\omega_4)$ is
the local 4-vertex. \cite{Hew93}. 
  The renormalised Hamiltonian (\ref{rham}) describes  only the low energy fixed point  and the leading irrelevant corrections.
 A complete description of
the original model can be obtained by working in the Lagrangian formulation.
The Lagrangian of the original model  ${\cal L}_{\rm
AM}(\epsilon_{d},\Delta,U)$ can be rewritten in terms of the renormalised
model
plus counter-terms, 
\begin{equation}
{\cal L}_{\rm AM}(\epsilon_{\mathrm{d}},
\Delta,U)={ \cal L}_{\rm AM}(\tilde\epsilon_{\mathrm{d}},
\tilde\Delta,\tilde U)+ {\cal L}_{\rm
  ct}(\lambda_1,\lambda_2,\lambda_3),\label{rlag}
\end{equation}
where ${\cal L}_{\rm
  ct}(\lambda_1,\lambda_2,\lambda_3)$ describes counter-terms.
In this formulation a renormalised perturbation expansion can be carried out in terms of the
quasiparticle propogaters in powers of the renormalised interaction
$\tilde U$. The role of the three parameters $\lambda_1$, $\lambda_2$
and $\lambda_3$ in the counter-term is to cancel any further
renormalisation of the parameters ,  $\tilde\epsilon_d$, $\tilde\Delta$ and
  $\tilde U$, which  are taken to be fully renormalised from the start \cite{Hew93, Hew01}.\\

In terms of
the renormalised parameters the exact results for the impurity spin and charge susceptibilities take a simple
form,
\begin{equation}
\chi_{s}={1\over 2}\tilde\rho(0)(1+
 \tilde U\tilde\rho(0)),\quad\chi_{c}={1\over 2}\tilde
 \rho(0)(1-\tilde
 U\tilde\rho(0)),\label{rsus}
\end{equation}
where  $\chi_s(0)$ is in units of  $(g\mu_{\rm B})^2$,
   $\chi_c(0)$ differs by a factor from the usual definition, so that for
the non-interacting system, $\chi_c(0)=\chi_s(0)$, and    
 $\tilde\rho(\omega)$ is the local density of states for the non-interacting quasiparticles, 
\begin{equation}
\tilde\rho(\omega)={\tilde\Delta/\pi\over
  (\omega-\tilde\epsilon_{d})^2+\tilde\Delta^2}.
\label{qpdos}
\end{equation} 

These results correspond to a first order calculation in the  renormalized
perturbation expansion in powers of  $\tilde U$ \cite{Yam75,Hew93}.\\

We can choose  to work solely in terms of the  parameters, $\tilde\epsilon_d$,
$\tilde\Delta$ and $\tilde U$,
and determine them by comparison with experiments, as one does in a
phenomenological Fermi liquid theory. To determine them in terms of the bare
parameters of the orginal model using equation (\ref{ren}) it would seem that
we need almost the complete solution of the model, including the dynamics. 
 However, they can be calculated  without recourse to these formulae by
 fitting the low energy excitations of the numerical renormalisation group
calculations to those of the effective low energy model given in equation 
(\ref{rham}), as was done essentially in the original NRG calculations \cite{Wil75,KWW80}.
The results for $\tilde\epsilon_d$,
$\tilde\Delta$ and $\tilde U$, for various parameter regimes of the model have
been described in earlier work \cite{HOM04}.
\\

\begin{figure}
\includegraphics[width=0.6\textwidth]{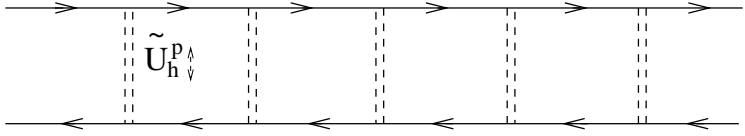}
\vskip.6cm
\caption{The repeated scattering of a quasiparticle $\uparrow$ and a
  quasihole $\downarrow$  which is taken into account in the RPT calculation
  of the dynamic spin susceptibility $\chi_s(\omega)$.}
\label{ph}
\end{figure}

We wish to generalise the results for the spin and charge susceptibilities given in
equation (\ref{rsus})  to obtain the corresponding dynamic susceptibilities:
the spin susceptibility, $\chi_s(\omega)$, which is given by the Fourier
transform of the double-time Green's functions of either  $\langle\langle d_{\uparrow}^{\dagger}(t)
d_{\downarrow}^{}(t); d_{\downarrow}^{\dagger}(t')
d_{\uparrow}^{}(t')\rangle\rangle/2$ (transverse susceptibility), or
$\langle\langle n_{{\rm d} \uparrow}(t)-n_{{\rm d}\downarrow}(t); n_{{\rm d}
  \uparrow}(t')-n_{{\rm d}\downarrow}(t')\rangle\rangle/4$ (paralllel susceptibility),
as they are equal in the absence of a magnetic field,
 and the charge susceptibility, $\chi_c(\omega)=\langle\langle n_{{\rm d} \uparrow}(t)+n_{{\rm d}\downarrow}(t): n_{{\rm d}
  \uparrow}(t')+n_{{\rm d}\downarrow}(t')\rangle\rangle/4$. To calculate these
 we
need to look at the processes involving repeated quasiparticle scattering.
A typical diagram in the renormalised perturbation expansion corresponding to
the repeated scattering of a spin up quasiparticle with a spin down quasihole 
is shown in figure \ref{ph}. These scattering 
terms at $\omega=0$ have implicitly been taken into account in the results for
the static susceptibilities in  (\ref{rsus}), and so we must introduce the
counter-term $\lambda_3$ to prevent overcounting. We 
define an irreducible 
interaction vertex $\tilde U^{p\uparrow}_{h\downarrow}=\tilde U-\lambda_3$,
  and sum the series
to derive an expression for the   dynamic spin susceptibility,
\begin{equation}
\chi_s(\omega)={1\over 2}{\tilde\Pi^{p\uparrow}_{h\downarrow}(\omega)\over 1-\tilde
  U^{p\uparrow}_{h\downarrow}\tilde\Pi^{p\uparrow}_{h\downarrow}(\omega)},\label{chisw}
\end{equation}
where 
$0.5\tilde\Pi^{p\uparrow}_{h\downarrow}(\omega)$ is the spin dynamic charge
susceptibility for the non-interacting quasiparticles. Expressions for
$\tilde\Pi^{p\uparrow}_{h\downarrow}(\omega)$
for the particle-hole symmetric model are given in the Appendix.
The approach is similar to the  random phase approximation (RPA) in terms of
quasiparticles \cite{SR83,KK00} and goes over to the RPA expression in the
weak coupling limit.
\\ 
\vskip1cm

\begin{figure}
\includegraphics[width=0.6\textwidth]{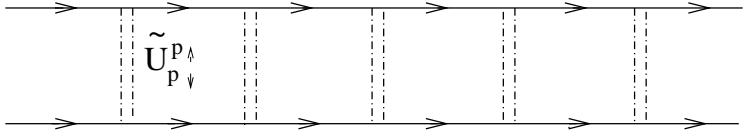}
\vskip.6cm
\caption{The repeated scattering of a quasiparticle  $\uparrow$ and a
quasiparticle  $\downarrow$ which is summed in the RPT calculation
  of the dynamic charge susceptibility $\chi_c(\omega)$.}
\label{pp}
\end{figure}

At zero frequency we can
identify
the result with the spin susceptibility given in equation (\ref{rsus}),
\begin{equation}
\chi_s(0)={1\over 2}\,{\tilde\rho(0)\over 1-\tilde
  U^{p\uparrow}_{h\downarrow}\tilde\rho(0)}={\tilde\rho(0)\over 2}(1+\tilde U\tilde\rho(0)), 
\end{equation} 
The effective particle-hole vertex can then be deduced in terms of  $\tilde U$,
\begin{equation}
 \tilde U^{p\uparrow}_{h\downarrow}={\tilde U\over 1+\tilde U\tilde\rho(0)}\label{uph}
\end{equation}
In terms of conventional Landau Fermi liquid notation,
$\tilde\rho(0)\tilde U^{p\uparrow}_{h\downarrow}=-F^a_0$. The resulting expression 
for  $\chi_s(\omega)$  in terms of  $\tilde U$ is
\begin{equation}
\chi_s(\omega)={1\over 2}\,{\tilde\Pi^{p\uparrow}_{h\downarrow}(\omega)(1+\tilde U\tilde\rho(0))\over
  1+\tilde U\tilde\rho(0)-\tilde
  U\tilde\Pi^{p\uparrow}_{h\downarrow}(\omega)}.\label{qrpas}
\end{equation}
This result is asymptotically exact as   $\omega\to 0$
as $\chi_s(\omega)\to \tilde\rho(0)(1+\tilde U\tilde\rho(0))/2$
(by construction), and its imaginary part satisfies the  Korringa-Shiba relation,
\begin{equation}
\lim_{\omega\to 0}{{\rm Im}\chi_s(\omega)
\over \pi\omega}={\tilde\rho^2(0)\over 2}(1+\tilde U\tilde\rho(0))^2=2\chi^2_s(0), 
\end{equation}
which is an exact relation
for the Anderson model \cite{Shi75}. We note that the result (\ref{qrpas}),
 in the large $U$
limit where $\tilde U\tilde\rho(0)\to 1$, is
equivalent to that derived  by Kuramoto and Miyake \cite{KM90}) based on a
semi-phenomenological Fermi liquid theory.
 \\
We repeat this exercise but this time consider the  
repeated scattering   two  opposite spin quasiparticles, as illustrated in
figure \ref{pp}, and 
define a corresponding irreducible 
interaction vertex $\tilde U^{p\uparrow}_{p\downarrow}$. 
The sum  of this series  gives for the  charge dynamic susceptibility,
\begin{equation}
\chi_c(\omega)={2\tilde\Pi^{p\uparrow}_{p\downarrow}(\omega)\over 1-\tilde
  U^{p\uparrow}_{p\downarrow}\tilde\Pi^{p\uparrow}_{h\downarrow}(\omega)}.
\end{equation}

\vskip1cm

Identifying the $\omega=0$ 
the result with the charge susceptibility in equation  (\ref{rsus}),
\begin{equation}
\chi_c(0)={1\over 2}\,{\tilde\rho(0)\over 1-\tilde
  U^{p\uparrow}_{p\downarrow}\tilde\rho(0)}={\tilde\rho(0)\over 2}(1-\tilde U\tilde\rho(0)).
\end{equation}
The effective particle-particle vertex can then be deduced in terms of  $\tilde U$,
\begin{equation}
 \tilde U^{p\uparrow}_{p\downarrow}={\tilde U\over 1-\tilde U\tilde\rho(0)}
\end{equation}
In terms of conventional Landau Fermi liquid notation,
$\tilde\rho(0)\tilde U^{p\uparrow}_{p\downarrow}=F^s_0$,
and the values of $F^s_0$ and $F^a_0$ are such as to satisfy the forward
scattering sum rule \cite{PN66},
\begin{equation}{F^s_0\over 1+F^s_0}+{F^a_0\over 1+F^a_0}=0.\end{equation}\par
\begin{figure}
\includegraphics[width=0.6\textwidth]{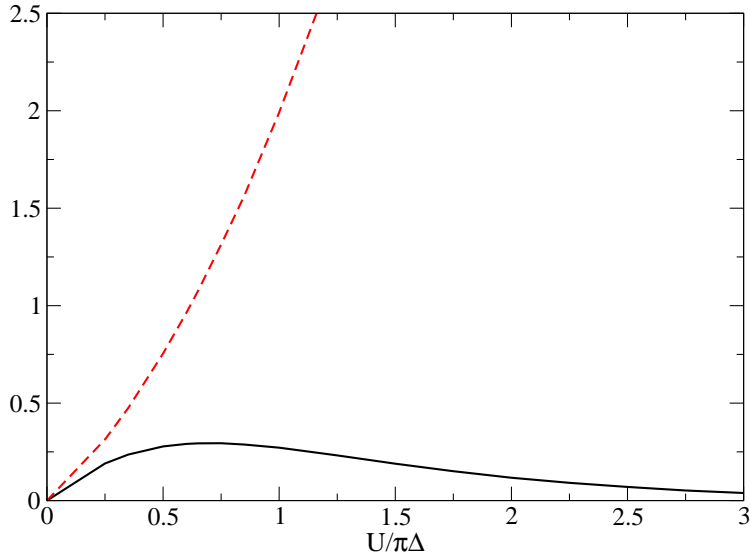}
\vskip.6cm
\caption{The irreducible vertices, $\tilde U^{p\uparrow}_{h\downarrow}(h)$
(full line) and
$\tilde U^{p\uparrow}_{p\downarrow}(h)$ (dashed line)
as a function of $U/\pi\Delta$
for the symmetric Anderson model.  }
\label{vertices}
\end{figure}

In figure \ref{vertices} we give the two irreducible vertices,
$\tilde U^{p\uparrow}_{h\downarrow}/\pi\Delta$ and $\tilde U^{p\uparrow}_{p\downarrow}/\pi\Delta$,
for the symmetric Anderson model as a function of $U/\pi\Delta$.
It can be seen that there is a rapid divergence between the interactions in the
spin and charge channels with increasing $U$. The vertex $\tilde U^{p\uparrow}_{h\downarrow}$
in the spin channel decreases monotonically with increase of $U$ and in the
limit
of large $U$ it tends to the value $2T_{\rm K}$, where $T_{\rm K}$ is the
Kondo temperature given by  $T_{\rm
 K}=\sqrt{(U\Delta/2)}e^{-\pi U/8\Delta+\pi\Delta/2U}$. The vertex
$\tilde U^{p\uparrow}_{p\downarrow}$
in the charge channel, on the other hand, increases monotonically though
remains finite and diverges only in the limit $U\to\infty$. \par

The RPT result for the dynamic charge susceptibility is
\begin{equation}
\chi_c(\omega)={1\over 2}\,{\tilde\Pi^{p\uparrow}_{h\downarrow}(\omega)(1-\tilde U\tilde\rho(0))\over
  1-\tilde U\tilde\rho(0)+\tilde
  U\tilde\Pi^{p\uparrow}_{h\downarrow}(\omega)},\label{qrpac}
\end{equation}
where       
$ \tilde\Pi^{p\uparrow}_{p\downarrow}(\omega)$ for the symmetric model
is given in the Appendix.
This result is again asymptotically exact and satisfies the relevant
Korringa-Shiba relation. In the next sections we evaluate these expressions
for the particle-hole symmetric model,
and compare the results with those calculated directly from numerical
renormalisation group calculations.
\par

\section{Dynamic Susceptibilities in the absence of a Magnetic Field}
 
We have shown that the RPT expressions for  $\chi_s(\omega)$ and
$\chi_c(\omega)$
are asymptotically exact in the limit $\omega\to 0$, but we would like to know
over what range of finite $\omega$ they give accurate results. To estimate
this we compare them with the corresponding results derived from a direct
numerical
renormalisation group (NRG) calculation for the dynamics \cite{SAK89,CHZ94}. These spectra can be
calculated using the same methods which have been developed for calculating
the spectral density of the one-electron impurity Green's function. These
direct calculations give a good overall picture of the behaviour over all
energy scales, but have some inaccuracies stemming mainly from 
the use of a discrete spectrum for the conduction electrons, and the logarithmic
broadening which is used to give the final continuous spectrum. To gauge
the accuracy of the direct NRG calculations we calculate $\chi_s(\omega)$
using this technique for the non-interacting system ($U=0$) and which we can
then compare with the exact result which for the particle-hole symmetric model
corresponds to equations (\ref{qrpas}) and (\ref{qrpac})
with $\tilde U=0$, $\tilde\epsilon_d=0$ and $\tilde\Delta=0$. We take
$\Delta=0.015708$
for these and all subsequent calculations.
\begin{figure}
\includegraphics[width=0.6\textwidth]{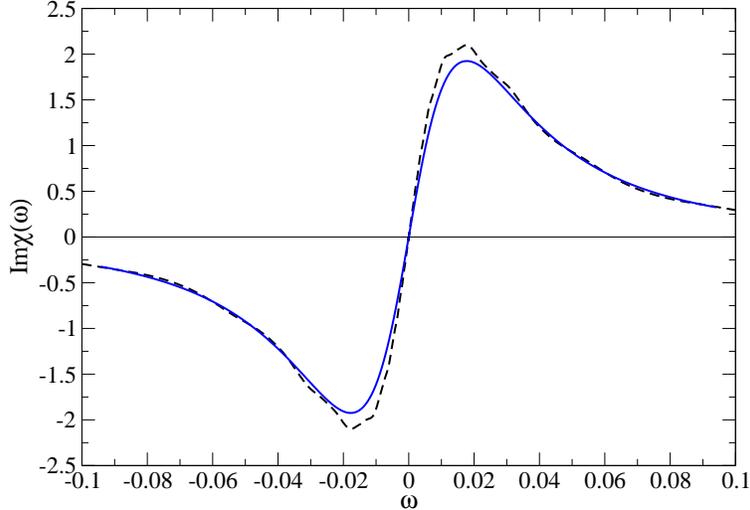}
\vskip.6cm
\caption{A comparison of the results of an NRG calculation
(dashed line) 
of the imaginary part of the spin susceptibility $\chi_s(\omega)$ for the non-interacting symmetric Anderson model
($U=0$) compared with an evaluation of the exact result (full line). }
\label{sym0.0}
\end{figure}

 The results
for  imaginary part of the retarded susceptibility, ${\rm Im}\,\chi_s(\omega)$ are shown in figure \ref{sym0.0}. We see that the direct 
NRG results give a reasonably good representation of the overall behaviour.
However, there is some overestimate of the gradient at $\omega=0$, leading
to a 10\% error in the Korringa-Shiba relation, and about a 5\% overestimate
of the peak height.\\

\begin{figure}
\includegraphics[width=0.6\textwidth]{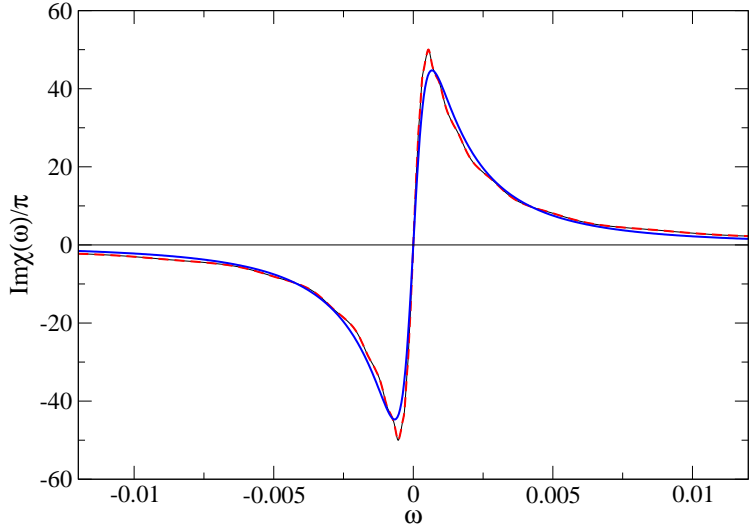}
\vskip.6cm
\caption{Results for the imaginary part of the dynamic spin susceptibility $\chi_s(\omega)/\pi$
for the symmetric model in the Kondo regime $U/\pi\Delta=3.0$.
The dashed curve corresponds to the NRG results and the full line to
the results using the RPT  expression, equation (\ref{qrpas}).}
\label{tsus3.0}
\end{figure}

\begin{figure}
\includegraphics[width=0.6\textwidth]{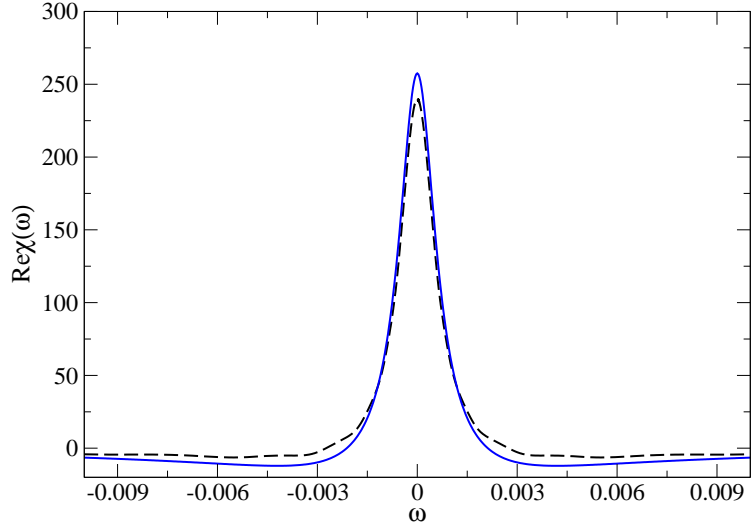}
\vskip.6cm
\caption{Results for the real part of the dynamic spin susceptibility $\chi_s(\omega)$
for the symmetric model in the Kondo regime $U/\pi\Delta=3.0$.
The dashed curve corresponds to the NRG results and the full line to
the results using the RPT expression, equation (\ref{qrpas}).}
\label{rchi3.0}
\end{figure}

We can now make a comparison of the direct NRG results in the interacting
case with the corresponding RPT results. We look at a strong correlation
situation first of all with $U/\pi\Delta =3.0$, in which the low energy charge excitations are almost
competely suppressed. The values of $\tilde U$ and $\tilde\Delta$ are
calculated
from the NRG levels as described in earlier work. The results for
${\rm Im}\,\chi_s(\omega)$ are shown in figure \ref{tsus3.0}. The peak structure is very much
enhanced compared with the non-interacting case and occurs on a scale
$\omega\sim\tilde\Delta=4T_{\rm K}/\pi$, where $T_{\rm K}$ is the Kondo
temperature. There is very good agreement between the two sets of results
over the whole range of this low-lying peak structure. It would appear that
the RPT results may be in fact more accurate than the direct NRG results,
as they satisfy the Korringa-Shiba relation exactly and, as we saw from the
results
in figure \ref{sym0.0},  the direct NRG results tend to overestimate the peak
heights.
The results for the real part of $\chi_s(\omega)$ is shown in figure
\ref{rchi3.0} and there is good agreement between the results but the direct
NRG underestimates the
value of  $\chi_s(0)$ by about 4\%.
 The corresponding dynamic charge susceptibilities are very small over
this range.
\\

\vskip1cm

\begin{figure}
\includegraphics[width=0.6\textwidth]{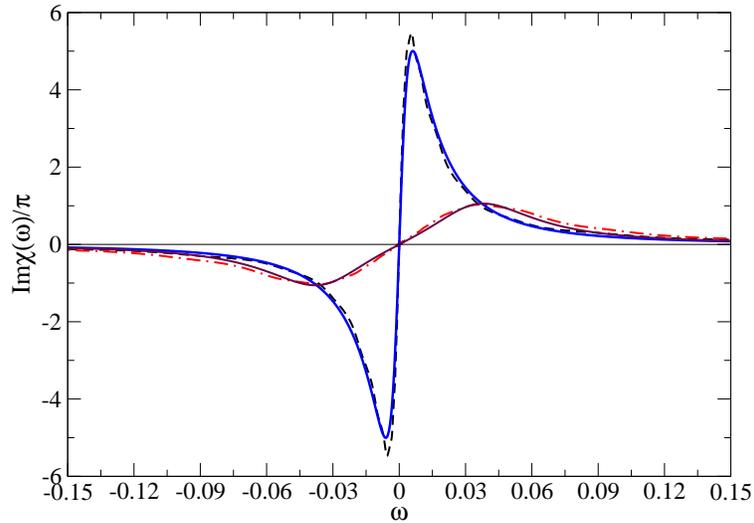}
\caption{Results for the imaginary parts of both the dynamic spin and charge susceptibilities susceptibility $\chi_s(\omega)$
for the symmetric model with $U/\pi\Delta=1.0$.
The dashed curve (spin) and the dot-dashed curve (charge) corresponds to the
NRG results and the full lines (thick/spin:thin/charge) to
the results using the RPT expressions.}
\label{comp1.0}
\end{figure}

We take a smaller value of $U$ such that $U/\pi\Delta=1.0$ to see  spin and
charge peaks of the same order. The NRG and RPT results for both ${\rm Im}\chi_s(\omega)$
and ${\rm Im}\chi_c(\omega)$ are shown in figure \ref{comp1.0}. The enhancement of the spin peaks
are down by a factor of 10 from the strong correlation case shown in figure \ref{tsus3.0},
and the $\omega$ range shown is greater by the same factor. In this moderately
correlated case the peaks in the charge spectrum are reduced but can be seen
on the same scale as the spin peaks. Again there is very good agreement
between the NRG and RPT results over this whole range, and the evidence from
the
earlier results suggest that the RPT results may be the more acccurate ones
over this range.\\
\vskip1cm

\begin{figure}
\includegraphics[width=0.6\textwidth]{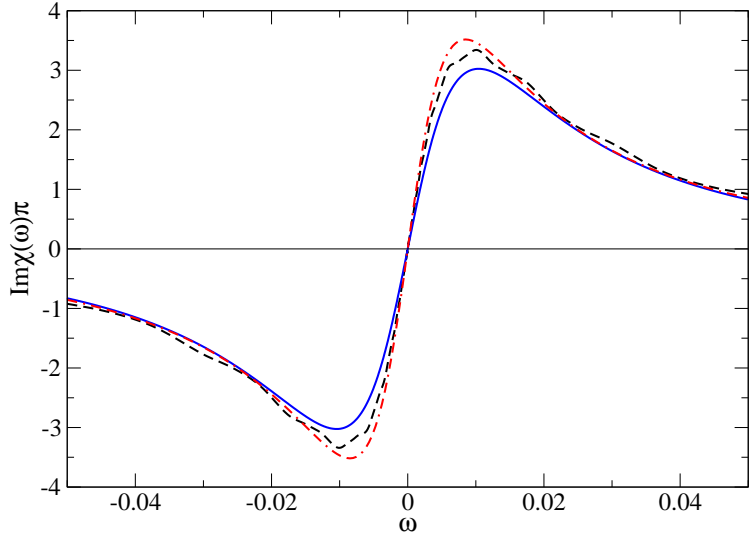}
\vskip.6cm
\caption{The imaginary part of the dynamic spin susceptibility for a weak
correlation case of the symmetic model with $U/\pi\Delta=0.5$.
The three sets of results are  
from (i) NRG calculations (dashed line), (ii) RPT (full line) and (iii) RPA
(dot-dashed line).}
\label{tsusimag0.5}
\end{figure}

We take a case of  weak correlation with $U/\pi\Delta=0.5$. In this case
it should be meaningful to make a comparision with the standard RPA results,
which correspond to taking the bare parameters in the RPT formulae;
$ U^{p\uparrow}_{h\downarrow}=U$, $\tilde\Delta=\Delta$. The results of all three calculations
for the imaginary part of the dynamic spin susceptibility are shown in  figure
\ref{tsusimag0.5}.  
All three curves  are qualitatively similar,
and there is very good agreement between them regarding the peak positions.
 The differences between the
NRG and RPT are similar to those noted in figures 4 and 6. If we compare the
peaks
with those in the non-interacting case shown in figure \ref{sym0.0}, we see that they are
enhanced only by a factor of the order 1.5. The peak heights are
overestimated in the RPA results. \\

\vskip1cm

\begin{figure}
\includegraphics[width=0.6\textwidth]{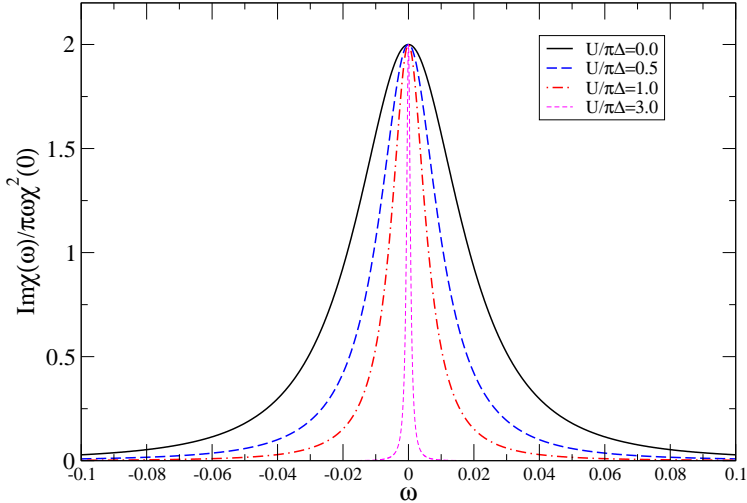}
\caption{Results for ${\rm Im}\chi_s(\omega)/\pi\chi_s^2(0)$ calculated
using the RPT for the symmetric model and a range of values of $U$.
}
\label{sratio}
\end{figure}

In figure \ref{sratio} we present the results for ${\rm Im}\,\chi_s(\omega)/\pi\chi^2_s(0)$,
for the four sets of results $U/\pi\Delta=0.0,0.5,1.0,3.0$. All sets have the
value
2 at $\omega=0$ corresponding to the Korringa-Shiba relation. With increasing 
$U$ there is the same narrowing to a scale set by the Kondo temperature as
in the one-electron spectral density of the impurity Green's function.
Combining the the Kramers-Kronig relation, $\int_{-\infty}^{\infty}{\rm
  Im}\chi(\omega)/\omega\, d\omega=\pi{\rm Re}\chi(0)$, with the Korringa-Shiba
relation gives
\begin{equation} {1\over \pi \chi^2_{s,c}(0)}\int_{-\infty}^{\infty}{{\rm
  Im}\,\chi_{s,c}(\omega)\over \omega}\, d\omega={1\over
  \chi_{s,c}(0)}={2\pi\tilde\Delta\over ( 1\pm\tilde
  U/\pi\tilde\Delta)}.\label{sumrule}\end{equation}
The total weight under the peak in figure \ref{sratio}, therefore,
when $U/\pi\Delta\gg 1$ and $\tilde U/\pi\tilde\Delta\to 1$,
is equal to  $4T_{\rm K}$. The sum rule (\ref{sumrule}) is satisfied precisely
  in the results shown. 

\begin{figure}
\includegraphics[width=0.6\textwidth]{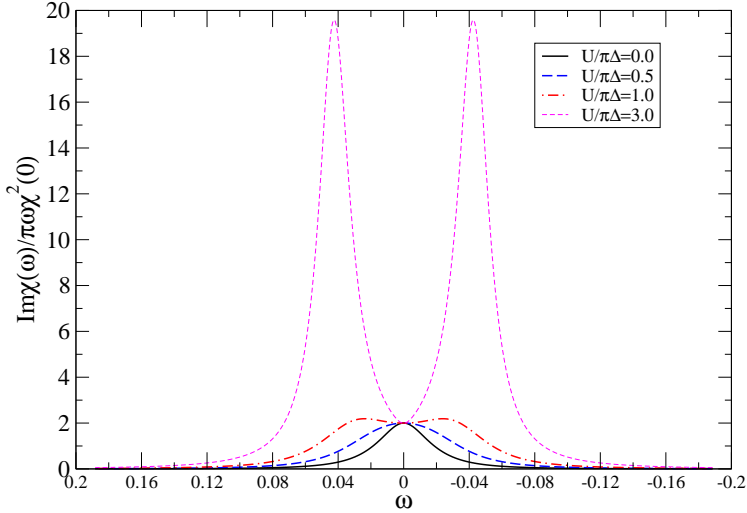}
\caption{Results for ${\rm Im}\chi_c(\omega)/\pi\chi_s^2(0)$ calculated
using the RPT for the symmetric model and the same  range of values of $U$
as for the spin susceptibility results shown in figure \ref{sratio}.}
\label{cratio}
\end{figure}

These results contrast to the corresponding results for the dynamic charge
susceptibility shown in figure \ref{cratio}. Again due to the  Korringa-Shiba relation all sets have the
value of 2 at $\omega=0$. The suppression of the charge fluctuations in this
case is indicated by the development of symmetric high energy peaks, with
values very much greater than the value at $\omega=0$.
From equation (\ref{sumrule}) it can be seen that as $U\to\infty$ and
$\tilde U/\pi\tilde\Delta\to 1$ that the area under the curves diverges
in this limit.
 However, though
qualitatively correct the RPT restricted to this class of diagrams cannot accurately describe the behaviour on this high energy scale.

\section{Dynamic Susceptibilities in a Magnetic Field}

\begin{figure}
\includegraphics[width=0.5\textwidth]{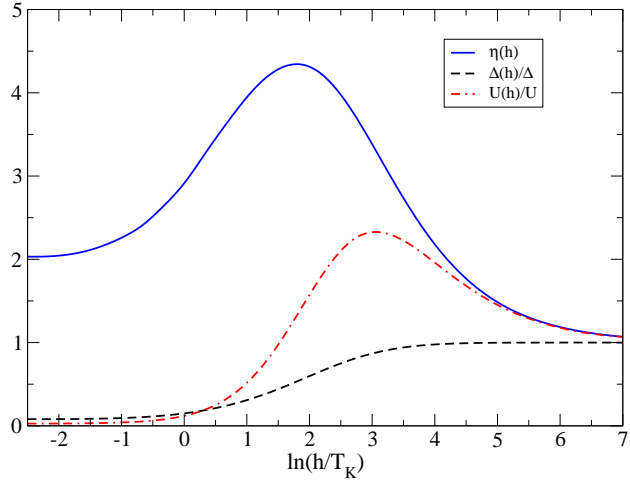}
\caption{The field dependence of the ratio of the renormalised parameters
relative to their bare values,
  $\eta(h)=\tilde h/h$,
 $\tilde\Delta(h)/\Delta$ and $\tilde U(h)/U$ for $U/\pi\Delta=3.0$ plotted as
a function of ${\rm ln}\,(h/T_{\rm K})$.}
\label{hparameters}
\end{figure}

We now consider the dynamic spin  response for the symmetric model in the presence of an applied magnetic field
 $H$. The low energy behaviour can still be described by a renormalised
Anderson model with parameters which now depend on the magnetic field
\cite{Hew05}.
The field dependent parameters 
$\tilde h$, $\tilde\Delta(h)$ and $\tilde U(h)$, are defined by
\begin{equation}\tilde h=z(h)(-{U\over2}+\Sigma(0,h)),\quad\tilde\Delta(h) =z(h)\Delta,\quad \tilde U(h)=z^2(h)\Gamma_{\uparrow\downarrow}(0,0,0,0:h),
\label{renh}\end{equation}
where $z(h)$ is given by
$z(h)={1/{(1-\Sigma'(0,h))}}$ and $h=g\mu_{\rm B}H/2$. These parameters can again be 
determined from the levels of a NRG calculation \cite{Hew05,HBK05}.
The induced impurity magnetisation $M_d(h)=(g\mu_{\rm B})^2m(h)$ at $T=0$,
 where $m(h)$
is given exactly by
 \begin{equation} m(h)={1\over \pi}{\rm tan}^{-1}\left({\tilde
 h(h)\over\tilde\Delta(h)}\right),\label{mag}
\end{equation}
a result which follows from the Friedel sum rule \cite{fsr}. The renormalized parameters, $\tilde
\eta(h)=\tilde h/h$, $\tilde\Delta(h)$ and $\tilde U(h)$, in the
presence of a magnetic field $H$, relative to their bare values, for
$U/\pi\Delta=3.0$ are given in figure \ref{hparameters}. The parameters are not all
independent and $\tilde U(h)$ can be deduced from the other two parameters
from the equation \cite{HBK05},
\begin{equation} 1+\tilde
  U(h)\tilde\rho(0,h)=\tilde\eta(h)+h{\partial\tilde\eta(h)\over\partial
  h}-{h\tilde\eta(h)\over\tilde\Delta(h)}{\partial\tilde\Delta(h)\over\partial
  h},\label{wi2}\end{equation}
where $\tilde\rho(\omega,h)$ is the quasiparticle density of states in the
presence of the magnetic field and is given by
\begin{equation}
\tilde\rho(\omega,h)={\tilde\Delta(h)/\pi\over
  (\omega-\tilde h(h))^2+\tilde\Delta^2(h)}.
\label{hdos}
\end{equation}
As the magnetic field increases the parameters change from  the Kondo regime
($h<T_{\rm K}$),
where $\tilde\eta(h)=2$, and $\tilde U(h)\tilde\rho(0,h)=1$, through a local
moment regime ($h>T_{\rm K})$, where $\tilde\eta(h)$ increases with $h$, and
$\tilde U(h)$ is enhanced to values greater than the bare value $U$, and
finally to the
almost free model when $h>U$, and the parameters revert to their bare values.\par

\vskip1cm

With a magnetic field present we have to distinguish between
the parallel and transverse spin susceptibilities,
$\chi_{s,\parallel}(\omega,h)$ and
$\chi_{s,\perp}(\omega,h)$. The static values at $T=0$ are given exactly by
\begin{equation}
\chi_{s,\parallel}(0,h)={1\over 2}\tilde\rho(0,h)(1+
 \tilde U(h)\tilde\rho(0,h)),\quad \chi_{s,\perp}(0,h)= {m(h)\over 2h}={1\over 2\pi h}{\rm tan}^{-1}\left({\tilde
 h(h)\over\tilde\Delta(h)}\right),\label{hsus}
\end{equation}

To calculate the frequency dependence of these  spin response functions
from the quasiparticle scattering  we  need to calculate the correponding irreducible particle-hole vertices
again as they will now depend on the value of the magnetic field. The
transverse susceptibility corresponding to a excitation with a change in the
$z$-component of the impurity spin of 1  involves the scattering of a
quasiparticle with spin up with a quasihole with spin down, and the
irreducible vertex in this channel we denote by $\tilde
U^{p\uparrow}_{h\downarrow}(h)$.
We derive its value using the same type of argument as earlier by requiring
the the sum of the series for $\omega=0$, should equal  the transverse
static susceptibility given in equation (\ref{hsus}).
We obtain the result,
\begin{equation}
 \tilde U^{p\uparrow}_{h\downarrow}(h)={\pi h(\tilde\eta(h)-1)\over{\rm tan}^{-1}
(\tilde h(h)/\tilde\Delta(h)) }.
\end{equation}
Using the results,
\begin{equation}
\lim_{h\to 0}(\tilde\eta(h)-1)=\tilde U\tilde \rho(0)={\tilde U\over \pi\tilde\Delta},\quad\quad \lim_{h\to 0}{{\rm tan}^{-1}
(\tilde h(h)/\tilde\Delta(h))\over h}=1+\tilde U\tilde \rho(0),
\end{equation}
it can be shown that  $\tilde
U^{p\uparrow}_{h\downarrow}(h)$, goes over to the earlier result in equation
(\ref{uph}) in the limit $h\to 0$. The resulting expression for $\chi_{s,\perp}(\omega,h)$
is as given in equation (\ref{qrpas}) with the $h$-dependence included in both
the scattering vertex $\tilde U^{p\uparrow}_{h\downarrow}(h)$ and $\tilde\Pi^{p\uparrow}_{h\downarrow}(\omega)$.\par

\begin{figure}
\includegraphics[width=0.6\textwidth]{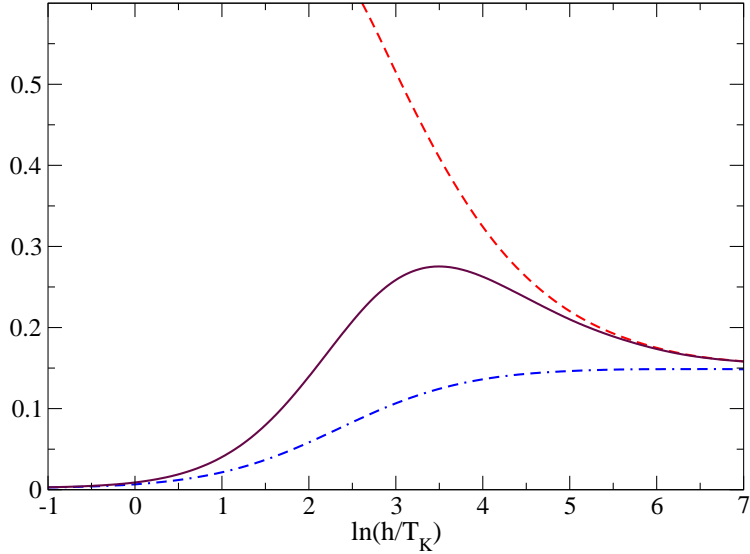}
\caption{The irreducible vertices, $\tilde U^{p\uparrow}_{h\downarrow}(h)$
(full line),
$\tilde U^{p\uparrow}_{h\uparrow}(h)$  (dot-dashed line), and
$\tilde U^{p\uparrow}_{p\downarrow}(h)$ (dashed line)
as a function of ${\rm ln}\,(h/T_{\rm K})$ for $U/\pi\Delta=3.0$. }
\label{hvertices}
\end{figure}

The parallel dynamic susceptibility corresponds to repeated scattering of a
quasiparticle with spin $\sigma$ and a quasihole with the same spin component $\sigma$, and hence no
change in the z-component of the spin. We denote the corresponding irreducible
vertex by $ \tilde U^{p\uparrow}_{h\uparrow}(h)$, corresponding to
$\sigma=\uparrow$, but for the symmetric model it is the same as for
$\sigma=\downarrow$. As the static parallel susceptibility has the same form
as earlier in equation (\ref{rsus}), apart for the $h$ dependence, the result is
\begin{equation}
 \tilde U^{p\uparrow}_{h\uparrow(p\downarrow)}(h)={\tilde U(h)\over (1\pm\tilde U(h)\tilde\rho(0,h))},
\end{equation}
with the corresponding result for the particle-particle vertex. Using the
results from figure \ref{hparameters}, we can deduce the three types of vertices
as a function of $h/\pi\Delta$ for $U/\pi\Delta=3.0$. The results
are shown in figure \ref{hvertices}. Initially for small $h$,  $ \tilde
U^{p\uparrow}_{h\downarrow}(h)=\tilde U^{p\uparrow}_{h\uparrow}(h)=2T_{\rm
  K}$, and then slowly diverge with $ \tilde
U^{p\uparrow}_{h\downarrow}(h)>\tilde U^{p\uparrow}_{h\uparrow}(h)$, and then slowly diverge with  but converge again for $h>U$ to the bare value   but converge again for $h>U$ to the bare value
$U$. The charge susceptibility, which is initially very large, decreases and
converges to the bare value $U$ for $h>U$.
\par

The transverse dynamic spin susceptibility 
is antisymmetric in the absence of a field,  ${\rm
 Im}\chi_{s,\perp}(\omega,0)=-{\rm Im}\chi_{s,\perp}(-\omega,0)$,
as can be seen in the results in figures \ref{sym0.0}-\ref{tsusimag0.5},
and as a consequence the total integrated  spectral weight is zero. This is no longer the
 case when $h\ne 0$, and the total spectral weight $w_\perp(h)$ is equal to
the average z-component of the impurity spin $m(h)=0.5\langle n_{d\uparrow}-
n_{d\downarrow}\rangle$. At $T=0$ the RPT result for
 $\chi_{s,\perp}(\omega,h)$ satisfies this result exactly as
\begin{equation}w_\perp(h)=\lim_{\omega\to\infty} \omega\chi_{s,\perp}(\omega,h)
=0.5\lim_{\omega\to\infty}\omega\tilde\Pi^{p\uparrow}_{h\downarrow}(\omega,h)
={1\over \pi}{\rm tan}^{-1}\left({\tilde h\over\tilde\Delta}\right),\end{equation}
which is equal to 
   $m(h)$ from equation (\ref{mag}).

 The total spectral weight $w_{\parallel}(h)$ for the parallel susceptibility
 $\chi_{s,\parallel}(\omega,h)$ on the other hand vanishes  as
\begin{equation}w_\parallel(h)=\lim_{\omega\to\infty} \omega\chi_{s,\parallel}(\omega,h)
=0.5\lim_{\omega\to\infty}\omega\tilde\Pi^{p\uparrow}_{h\uparrow}(\omega,h)
=0.\end{equation} 

 \begin{figure}
\includegraphics[width=0.5\textwidth]{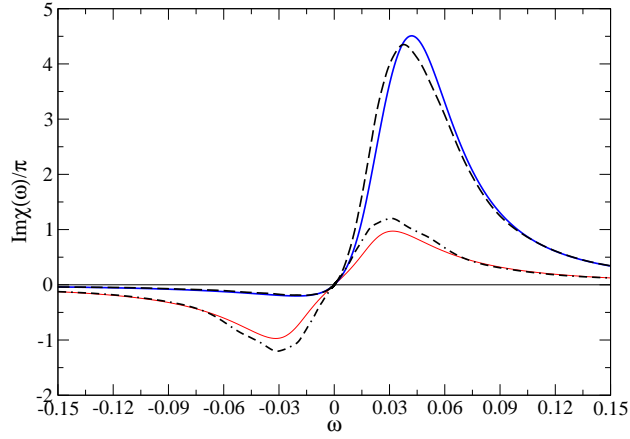}
\caption{The NRG results for the imaginary parts of $\chi_{s,\perp}(\omega,h)$
 (dashed line) and
$\chi_{s,\parallel}(\omega,h)$ (dot-dashed)
as a function of $\omega$ for $U=0$ compared with the exact results
($\perp$-thick full line,
$\parallel$-thin full line) for $h=0.4\pi\Delta$.}
\label{chis.4del}
\end{figure}
  
\begin{figure}
\includegraphics[width=0.5\textwidth]{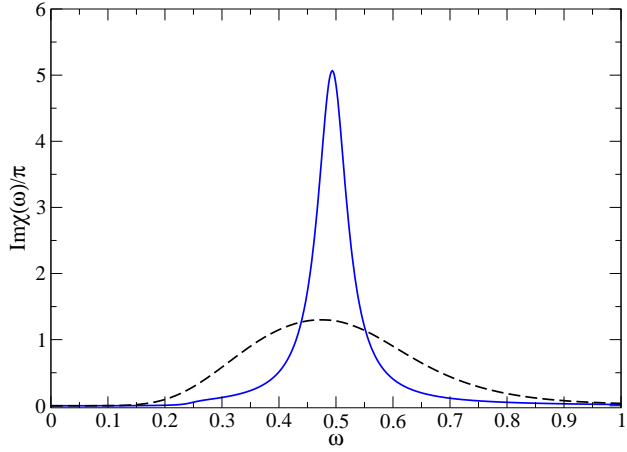}
\caption{The NRG results for the imaginary parts of $\chi_{s,\perp}(\omega,h)$
 (dashed line) and
as a function of $\omega$ for $U=0$ compared with the exact results (full
line) for $h=5\pi\Delta$.}
\label{chit5del}
\end{figure}

To gauge the errors in NRG results \cite{note} due to the use of a discrete spectrum and
logarithmic broadening we again compare them with  exact results for $U=0$.
The results of this comparison for $h=0.4\Delta$ are shown in figure \ref{chis.4del}.
 We see that the NRG results give the correct overall forms for the imaginary
 parts for both the
 transverse and parallel susceptibilities.
 However, due to the discrete
 spectrum and broadening on this relatively high energy scale, there is
there is some deviation from the exact results. The asymmetry of the results
for the transverse case due to the field is
apparent.    In figure \ref{chit5del} we compare
 the NRG results for the imaginary part of the transverse susceptibility with
 the
exact results for an extremely large magnetic field $h=5\Delta$. On this 
scale the impurity is effectively an almost localised and nearly fully
polarised
local moment. The spectrum, therefore, consists of a single peak at the energy
required to flip the spin in the applied field, $\omega_p\sim 2h$. Whereas the
exact result has a sharp peak the NRG is much broader on this high energy
scale,
but the peak position and total spectral weight of the NRG result are
correct. \par

\begin{figure}
\includegraphics[width=0.5\textwidth]{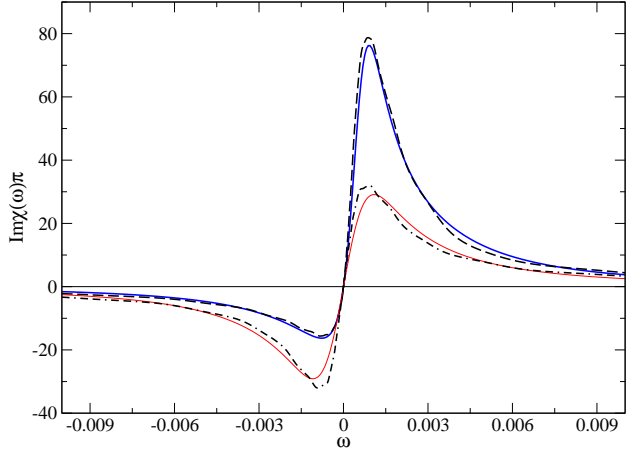}
\caption{The RPT results for the imaginary parts of $\chi_{s,\perp}(\omega,h)$
 (thick full line) and
$\chi_{s,\parallel}(\omega,h)$ (thin full line) for $h=0.4T_{\rm K}$
as a function of $\omega$ for $U/\pi\Delta=3.0$ compared with the NRG results
($\perp$-dashed line,
$\parallel$-dot-dashed line).}
\label{chis.4tk}
\end{figure}

We now compare the NRG results with those of the RPT results in the strong
coupling
case corresponding to the Kondo limit. In figure \ref{chis.4tk} we compare the imaginary
parts of the parallel and transverse spin susceptibilities
for a magnetic field $h=0.4T_{\rm K}$. We see that there is good agreement
between the two sets of results. Comparing the differences with those seen
in figure \ref{chis.4del}, where the NRG results are compared with the exact ones, one
would
conclude that the RPT results are probably the more accurate ones over this
range.

 The question arises as to whether the RPT results can give similarly
good results for higher magnetic field values. As we increase the magnetic
field the Kondo effect is suppressed and for extremely large fields the
renormalised parameters revert to their bare values as can be seen in 
the results shown in figure \ref{hparameters}. We increase the applied field to test
the results over this range of fields. We concentrate on the transverse
susceptibility
as this can change significantly with magnetic field as we pass from the Kondo 
regime to the local moment regime, while the parallel susceptibility mainly
gets suppressed with increase of field.

\begin{figure}
\includegraphics[width=0.5\textwidth]{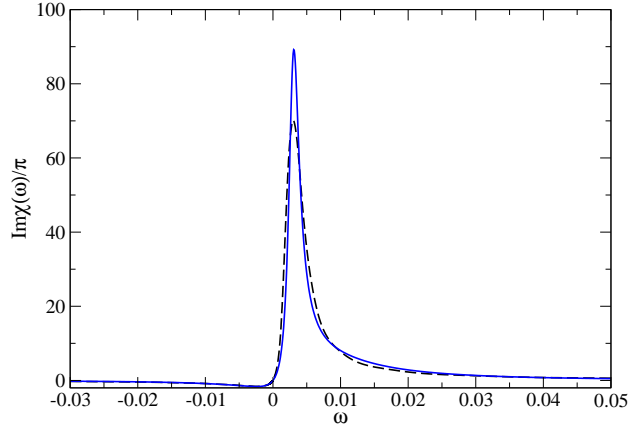}
\caption{The RPT results for the imaginary parts of $\chi_{s,\perp}(\omega,h)$
 (thick full line)  for $h=2T_{\rm K}$
as a function of $\omega$ for $U/\pi\Delta=3.0$ compared with the NRG results
(dashed line).}
\label{chit2tk}
\end{figure}

In figure \ref{chit2tk} we show the imaginary part of the transverse susceptibility
for $U/\pi\Delta=3.0$ for $h=2T_{\rm K}$
(${\rm ln}\,h/T_{\rm K}=0.693$). There is quite a dramatic change from
the results for $h=0.4T_{\rm K}$, shown in figure \ref{chis.4tk}. The spectral
weight in the region $\omega<0$ is almost completely suppressed,
and there is a single sharp peak in the spectrum for $\omega>0$.
 Again there is remarkable agreement of the RPT
results with those obtained from the direct NRG calculations.\par
\begin{figure}
\includegraphics[width=0.5\textwidth]{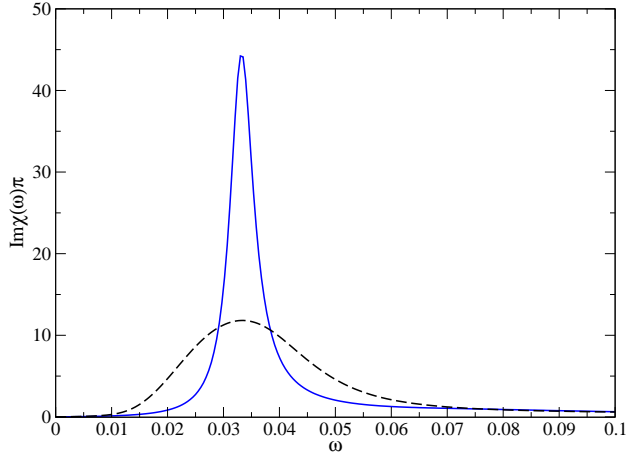}
\caption{The RPT results for the imaginary parts of $\chi_{s,\perp}(\omega,h)$
 (thick full line) for $h=20T_{\rm K}$
as a function of $\omega$ for $U/\pi\Delta=3.0$ compared with the NRG results
(dashed line).}
\label{chit20tk}
\end{figure}

For values of the magnetic field such that ${\rm ln}(h/T_{\rm K})\sim 3.0$,
we see from figure \ref{hparameters} that $\tilde U(h)$ is enhanced over its
bare value $U$, corresponding to the local moment regime. It is
interesting to see whether the Fermi liquid theory can be applied also
in this regime. In figure \ref{chit20tk} we plot the transverse spin spectral density
calculated with the NRG and RPT for $h/T_{\rm K}=20$ (${\rm ln}\,h/T_{\rm K}=3.0$). We see that remarkably that there is good
agreement in this regime as well, both the peak positions and total spectral
weight are in agreement. The peak position corresponds to $0.86\times 2h$
 Though it can be seen from figure \ref{hparameters} that the renormalised
field parameter $\tilde h(h)$ is
enhanced over $h$, the peak position is not at $2\tilde h$, and is somewhat
less than the simple Zeeman splitting $2h$. The NRG results are much broader but this
is because a discrete spectrum had to be used, and the broadening, which is
applied on a logarithmic scale, is much larger at higher values of $\omega$.

\begin{figure}
\includegraphics[width=0.5\textwidth]{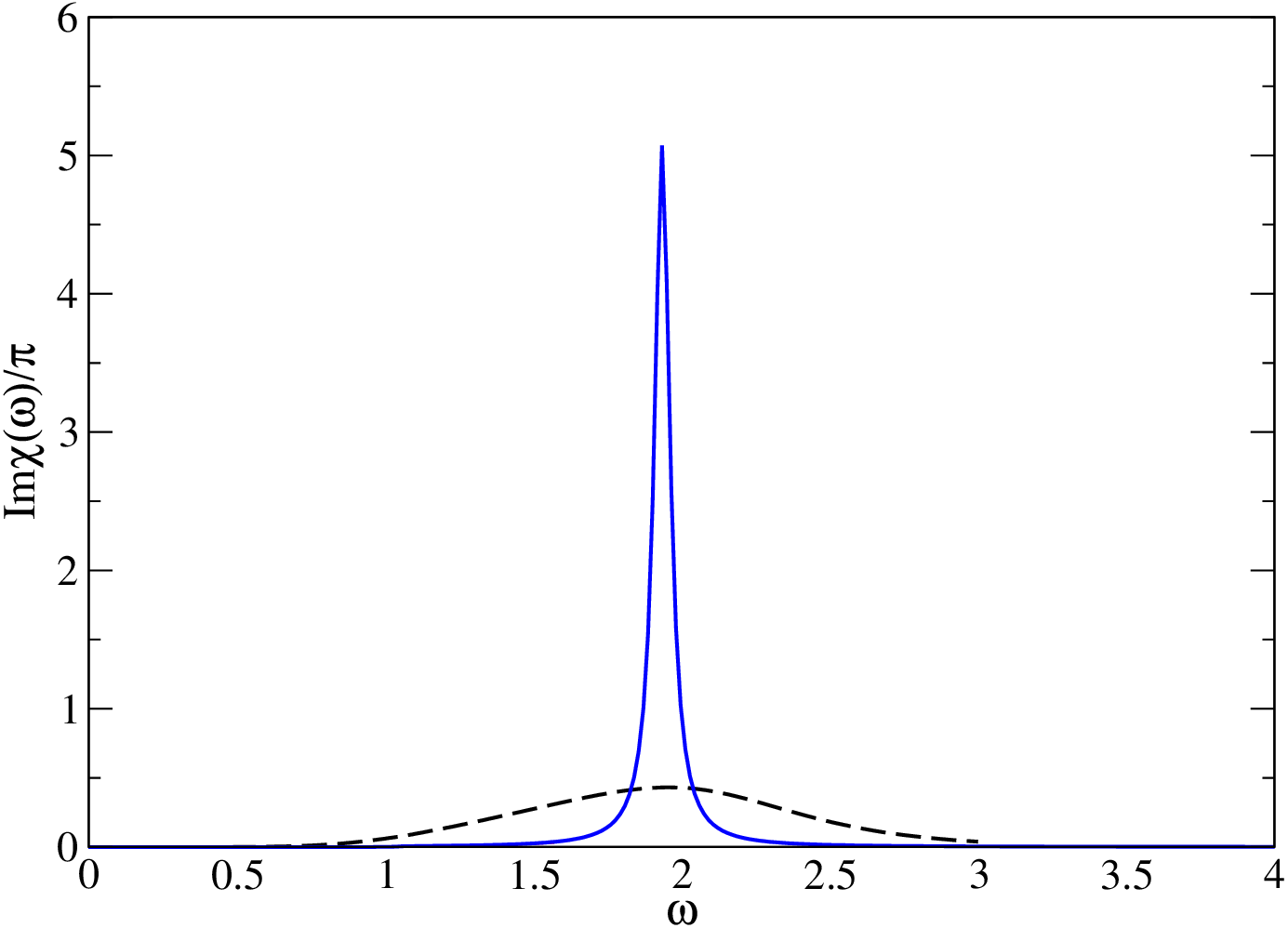}
\caption{The RPT results  for the imaginary parts of $\chi_{s,\perp}(\omega,h)$
 (thick full line) for $h=10^3T_{\rm K}$
as a function of $\omega$ for $U/\pi\Delta=3.0$ compared with the NRG results
(dashed line).}
\label{chit1000tk}
\end{figure}

Finally, we check the extreme limit, $h/T_{\rm K}=10^3$ (${\rm ln}\,h/T_{\rm
  K}=6.93$), where it can be seen in figure \ref{hparameters} that the
renormalised parameters revert to their bare values. Again from
the results given in figure \ref{chit1000tk} it can be seen that the two sets of results correspond. 
In this regime the peak is at $\omega_p =2h$, and all many-body
effects  have disappeared.\par

\begin{figure}
\includegraphics[width=0.5\textwidth]{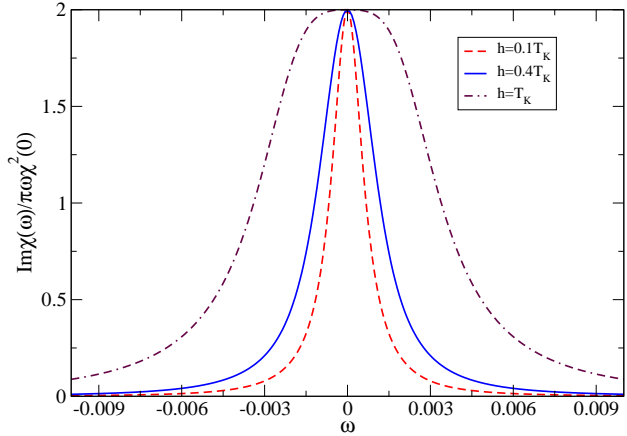}
\caption{Results for ${\rm Im}\chi_{s,\parallel}(\omega,h)/\pi\chi_{s,\parallel}^2(0,h)$ calculated
using the RPT equations for a range of values of the magnetic field  $h$.}
\label{pratio}

\end{figure}

The similarity of the parallel susceptibility to the susceptibilities
in the absence of a magnetic field suggests that the this response
function should satisfy the Korringa-Shiba relation. We test this out
 by plotting the imaginary value of $\chi_\perp(\omega,h)$ divided by
$\omega\pi\chi^2_\perp(0,h)$ in figure \ref{pratio}  for $U/\pi\Delta=3.0$ and
$h/T_{\rm K}=0.1,0.4,1$ (the parallel suceptibility is very much suppressed
 for larger values of $h$). We see the the  Korringa-Shiba relation
does apply and with increasing magnetic field the width of the central peak
broadens, reflecting the suppression of the Kondo resonance with the
increasing values of the
magnetic field. The increase of area under the  peak with increase magnetic field, from
equation  (\ref{sumrule}), can be related to the corresponding decrease in the
static parallel susceptibility. \par

\section{Conclusions}
We have taken  into account  processes involving repeated
quasiparticle scattering to calculate the  dynamical
charge and spin susceptibilities at zero temperature.
 As we use the exact irreducible vertices at $\omega=0$,
 we know from  Fermi liquid theory  that the results will be asymptotically exact 
as $\omega\to 0$, and that the Korringa-Shiba relation will be satisfied. We
have also shown that the total spectral  weight is given correctly. However,
it is not self-evident that taking into account these processes alone in the 
RPT   will constitute a good approximation over all the relevant
 the energy
scales. By comparing the results with those obtained by a direct evaluation
of the response functions using the NRG we have established that the results
of the RPT calculations give a remarkably accurate description of the 
dynamic spin susceptibilities for the Anderson model with $U>0$ over all
the relevant energy scales and arbitrary values of the external magnetic field.
The earlier work of Kuramoto and Miyake \cite{KM90} came to the same
conclusion, though their derivation is restricted to the Kondo limit, and
with no magnetic field.\par
 Whereas the spin excitations are enhanced and pushed to lower frequencies
 for large values of $U$,  the low energy charge
excitations are suppressed and pushed to higher energies.
  The RPT
results in this case can only accurately describe the low energy
suppression, as the high energy peaks in the charge spectrum are pushed
to energies of the order $\omega\sim U/2$, where the low energy renormalised
parameters used are not appropriate. It is the spin fluctations, however, which are the important fluctuations in this regime, which are enhanced
and pushed to lower energies. The energy scale of these fluctuations
in the large $U$ regime is the Kondo temperature $T_{\rm K}$, which is the 
renormalised energy scale, and over this scale the RPT is an excellent
approximation. For the negative $U$ Anderson model the opposite will be the
case. For large absolute values of $U$
 the spin fluctuations are suppressed and the charge fluctuations enhanced
and driven to low energies. In this case it will be the charge fluctuations
that are physically the more important ones, and for these the RPT will
provide a good description.\par
In calculating the quasiparticle contribution to the low energy spectral density of the
single particle Green's function $ G_{\sigma}(\omega)$, it has to be
multiplied
by the quasiparticle weight factor $z$ (wave function renormalisation factor).
At first sight it might seem surprising  that the quasiparticle contribution
to the dynamic susceptibility does not have to be multiplied by a factor
$z^2$. The reason why such a factor is not required is because there is also
a renormalisation of the vertex $\Lambda(\omega_1:\omega)$ which couples the system to the external magnetic
field. It can be shown via a Ward identity that this vertex  evaluated at
$\omega_1=\omega=0$ is equal  to $1/z$ \cite{Noz64} and, as a consequence, on
differentiating
twice with respect to $H$ brings in a factor $1/z^2$ which cancels 
factors of $z$ from product of the full Green's functions. As a result
the $z$ does not appear in calculating the dynamical response functions
from the RPT.\par
The approach used here can be generalised to lattice models, such as the
Hubbard model and Hubbard-Holstein models,  within the
dynamical mean field theory, and these calculations are in progress.

\bigskip
\noindent{\bf Acknowledgement}\par
\bigskip

  We thank the EPSRC for support through the Grant GR/S18571/01),
and  Winfried Koller, Dietrich Meyer and Johannes Bauer for helpful discussions
and for contributions to the development of the NRG programs. We also thank
Jim Freericks for comments on an earlier draft of the paper.

\par
\section{Appendix}
The  free local  quasiparticle propagator $\tilde G_{\sigma}(\omega)$
for the symmetric Anderson model, with a magnetic field $H$ acting at the
impurity site, 
is given by 
\begin{equation}\tilde G_{d\sigma}(\omega)={1\over\omega+i\tilde\Delta{\rm
      sgn}(\omega)+\sigma\tilde h}\end{equation}
in the zero temperature perturbation theory formalism 
There are two indepndent quasiparticle pair  propagators defined by 
\begin{equation}
\tilde\Pi^{p\sigma}_{h\sigma'}(\omega,\tilde
h)={i\over 2\pi}\int_{-\infty}^{\infty}\tilde G_{\sigma}(\omega+\omega',\tilde h)
\tilde G_{\sigma'}(\omega',\tilde h){d\omega'},\end{equation}
and
\begin{equation}
\tilde\Pi^{p\sigma}_{p\sigma'}(\omega,\tilde
h)={i\over 2\pi}\int_{-\infty}^{\infty}\tilde G_{\sigma}(\omega-\omega',\tilde h) 
\tilde G_{\sigma'}(\omega',\tilde h){d\omega'}.\end{equation}
These integrals can be evaluated analytically and the results for $\omega>0$ are 
$$\tilde\Pi^{p\uparrow}_{h\uparrow}(\omega,\tilde h)={\tilde\Delta\over{\pi(\tilde h^2+\tilde\Delta^2)}}\quad{\rm for}\quad\omega= 0,$$
\begin{equation}={-\tilde\Delta\over{\pi\omega(\omega+2i\tilde\Delta)}}\left\{{\rm
      ln}\left({{\omega+i\tilde\Delta-\tilde h}\over{i\tilde\Delta-\tilde h}}\right)+
{\rm ln}\left({{\omega+i\tilde\Delta+\tilde h}\over{i\tilde\Delta+\tilde h}}\right)\right\}\quad{\rm for}\quad\omega\ne 0.\end{equation}
and
$$\tilde\Pi^{p\uparrow}_{h\downarrow}(\omega,\tilde h)={i\over{\pi(i\tilde\Delta-\tilde h)}} -
{1\over 2\pi\tilde \Delta}{\rm
  ln}\left({{i\tilde \Delta-\tilde h}\over{i\tilde\Delta+\tilde
  h}}\right)\quad{\rm for}\quad\omega=
  -2\tilde h,$$
\begin{equation}={-i\over{\pi}}\left\{{1\over{\omega+2\tilde
  h+2i\tilde\Delta}}{\rm ln}\left({{\omega+i\tilde\Delta+\tilde
    h}\over{i\tilde\Delta+\tilde h}}\right)-{1\over{\omega+2\tilde h}}
{\rm ln}\left({{\omega+i\tilde\Delta+\tilde h}\over{i\tilde\Delta-\tilde
    h}}\right)\right\}\quad{\rm for}\quad\omega\ne -2\tilde h.\end{equation}

\noindent The formulae for $\omega<0$ are obtained by replacing $i$ with $-i$
in the above. For particle-hole symmetry, we have $
\tilde\Pi^{p\uparrow}_{p\downarrow}(\omega,\tilde
h)=-\tilde\Pi^{p\downarrow}_{h\downarrow}(\omega,\tilde h)$,
and 
$\tilde\Pi^{p\downarrow}_{h\downarrow}(\omega,\tilde
h)=\tilde\Pi^{p\uparrow}_{h\uparrow}(\omega,\tilde
h)=\tilde\Pi^{p\downarrow}_{h\downarrow}(\omega,-\tilde h)$


\begin{thebibliography}{99}
\bibitem{note} In the NRG calculations in the presence of a magnetic field we
adopt the density matrix modified NRG approach introduced by Hofstetter \cite{Hof00}.
\end{thebibliography}
\end{document}